\documentclass[aps,prd,tightenlines,preprintnumbers,showpacs,superscriptaddress,nofootinbib,eqsecnum,floatfix,amsmath,longbibliography,
twocolumn]{revtex4-2}
\usepackage{bm}
\usepackage{graphicx}
\usepackage[pdftex]{color}

\usepackage[dvipsnames]{xcolor}

\definecolor{dblue}{rgb}{0.0, 0.0, 0.6}
\definecolor{ddblue}{rgb}{0.0, 0.0, 0.4}
\definecolor{dgreen}{rgb}{0.0, 0.4, 0.0}
\definecolor{dgray}{gray}{0.4}
\definecolor{ddgray}{gray}{0.3}

\usepackage{hyperref}

\def\floatcaption#1#2{ \caption{#2 \label{#1}}}

\def\m{\mu}

\def\cbo{{\,\raise-.15ex\Sc [\,}}                       

\def\svev#1{\left\langle #1\right\rangle}       

\def\ddt#1{{\buildrel {\hbox{\LARGE .\kern-2pt.}} \over {#1}}}

\def\ie{\mbox{\it i.e., }}

\def\Tr{{\rm Tr}\,}

\def\Det{{\rm Det}}

\def\Eq#1{Eq.~(\ref{#1})}

\long \def \blockcomment #1\endcomment{}

\def\be{\begin{equation}}
\def\ee{\end{equation}}
\def\bea{\begin{eqnarray}}
\def\eea{\end{eqnarray}}

\def\mPV{m_{\text{PV}}}
\def\MPV{M_{\text{PV}}}
\def\NPV{N_{\text{PV}}}
\def\Seff{S_{\text{eff}}}
\def\Sind{S_{\text{ind}}}
\def\gGF{g^2_{\text{GF}}}
\def\mf{m}

\def\vs{{\em vs.}}
\def\svev#1{\left\langle #1\right\rangle}

\begin{document}

\title{Taming lattice artifacts with Pauli--Villars fields}

\author{Anna Hasenfratz}
\affiliation{Department of Physics, University of Colorado, Boulder, CO 80309, USA}

\author{Yigal Shamir}
\affiliation{Raymond and Beverly Sackler School of Physics and Astronomy,
Tel~Aviv University, 69978 Tel~Aviv, Israel}

\author{Benjamin Svetitsky}
\affiliation{Raymond and Beverly Sackler School of Physics and Astronomy,
Tel~Aviv University, 69978 Tel~Aviv, Israel}


\begin{abstract}
As fermions are added to a lattice gauge theory, one is driven to stronger bare coupling in order to maintain the same renormalized coupling.
Stronger bare couplings are usually associated with larger gauge fluctuations, leading to larger cutoff effects and more expensive simulations.
In theories with many light fermions,
sometimes the desired physical region
cannot be reached before encountering a phase boundary.
We show that these undesired effects can be reduced
by adding Pauli--Villars fields. We reach significantly larger
renormalized couplings while at the same time damping
short-distance fluctuations of the gauge field.  This may allow for
controlled continuum extrapolations from large lattice spacings.
\end{abstract}

\maketitle

\newpage
\section{\label{intro} Introduction}

There is wide interest in the non-perturbative properties
of asymptotically free gauge theories at strong renormalized coupling.
Lattice gauge theory simulations aim to investigate strong-coupling physics
at large distances while starting at weak coupling
at short distances. Tuning the bare coupling to zero
decreases the lattice spacing and reduces cutoff effects
but requires increasing the lattice volume
to keep the physical volume sufficiently large.
Various improvement programs, both perturbative and non-perturbative,
attempt to reduce lattice artifacts
so that simulations even at reasonable lattice volumes
can predict continuum physics with high precision.
This approach works well when the gauge coupling runs fast,
as in QCD with a few flavors of light fermions.

As is well known, fermion fields screen the gauge interaction
and thus slow down the running of the gauge coupling.
Sufficiently many fermions even lead to the emergence of
an infrared fixed point. In slowly running systems,
computationally acceptable lattice volumes do not allow significant change
from the bare to the renormalized coupling. In order to reach
strong renormalized couplings, simulations must start
with a large bare coupling. This, in turn, brings in large cutoff effects.
Often the large ultraviolet fluctuations induce
unwanted lattice artifacts like bulk phase transitions.

We illustrate the problem with Fig.~\ref{gsq-beta-nhyp-noPV}.
The top panel shows $\gGF$, the gradient-flow coupling\footnote{
  For the precise definition of the gradient-flow coupling
  see Sec.~\ref{action}.
}
at fixed lattice distance as a function of the bare coupling $\beta$
for SU(3) gauge theory coupled to nHYP staggered fermions
with $N_f=0$, 4, 8, and 12 (continuum) flavors.
Each theory except the pure Yang--Mills theory
exhibits a first-order bulk transition to an S4 phase\footnote{
  The S4 phase is characterized by the spontaneous breaking
  of the staggered shift symmetries \cite{Cheng:2011ic,Hasenfratz:2013uha,Kotov:2021mgp}. For a possible explanation  of this symmetry breaking pattern see \cite{Aubin:2015dgk}.
}
at strong coupling.
The leftmost point in each dataset is slightly to the right
of the phase transition for that model.
One can see that the largest $\gGF$ that can be reached before
encountering the phase transition gets smaller with increasing $N_f$.
The bottom panel of Fig.~\ref{gsq-beta-nhyp-noPV} shows the corresponding
plaquette values (normalized such that the maximal value is 3).
Comparing the two panels reveals that, as the number of fermions increases,
reaching the same value of $\gGF$ (if possible) is accompanied by
a smaller plaquette value. In other words, it comes at the price
of stronger short-distance gauge field fluctuations.

Small plaquette values typically signal a difficult simulation.
Worse is the presence of the phase boundary.
There is the possibility that simulations intended to study the properties
of the Gaussian fixed point at $g_0^2=0$ but performed near
the first-order transition will reflect the properties of the bulk transition.
In any case, bulk phase transitions limit the physical range
that lattice simulations can probe.

\begin{figure}[t]
\vspace*{-1ex}
\includegraphics[width=0.99\columnwidth]{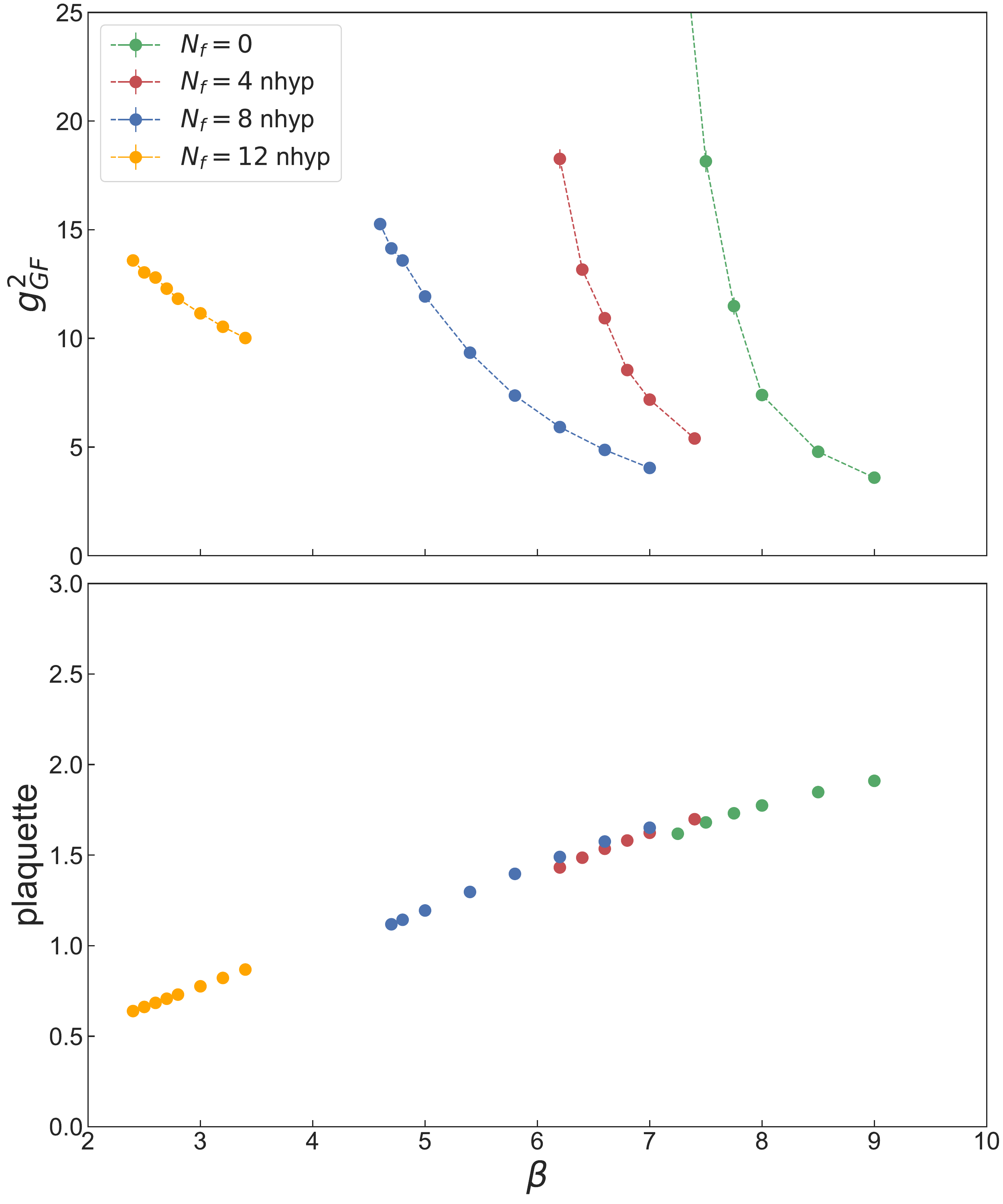}
\vspace*{-2ex}
\floatcaption{gsq-beta-nhyp-noPV}%
{Gradient flow coupling $\gGF$ (top)
and plaquette (bottom) \vs\ the bare coupling $\beta$
for SU(3) gauge theory with
$N_f=0$, 4, 8, and~12 flavor staggered fermions,
with nHYP smearing,  on lattice volume $8^4$. The finite-volume gradient-flow coupling of \Eq{GF} is defined by the choice $c=0.45$. The fermion mass is $a\mf=0.005$.
All systems but the pure Yang--Mills theory show a first order transition
to an $S^4$ phase, just to the left of the corresponding data set.
}
\end{figure}

Cutoff effects will be reduced
if we can suppress ultraviolet fluctuations
without affecting the infrared physics.
In this paper we show that adding to the lattice action
a set of heavy Pauli--Villars (PV) fields achieves this goal.
These are bosonic Dirac fields that we choose to have the same
lattice action as the fermions,
but with heavy masses fixed at $a\mPV=\mathcal{O}(1)$.
These PV fields play a role similar to that of the continuum PV fields
in regularizing fermions at large momenta \cite{Pauli:1949zm}.
While fermions screen the gauge coupling, the PV fields anti-screen.
At short distances $r\alt 1/\mPV$ they will enhance the running of the coupling.
At large distances $r\gg a$ the running of the gauge coupling is physical,
and will not be altered by the presence of the PV fields.

Matter fields whose mass is fixed in lattice units,
whether fermions or bosons, decouple in the continuum limit where the
lattice spacing $a\to 0$ or the lattice correlation length $\xi \to \infty$.
The effect of the PV bosons can be summarized as an addition
to the lattice action.  They generate a local effective gauge action \cite{Appelquist:1974tg}
that can balance the screening effect of the fermions at short distances.
This recalls the old idea of  gauge fields wholly generated
by heavy matter fields \cite{Hasenfratz:1992jv,Kazakov:1992ym},
but here the goal is only to compensate for the
short-distance effects of lattice fermions.
While continuum PV fields are introduced in one-to-one correspondence
with the fermions in order to cancel UV divergences,
on the lattice we may allow the number of PV fields to be
a free parameter. The induced term in the gauge action is simply proportional
to the number of PV fields.%
\footnote{PV bosons are also included when using lattice domain wall fermions, in one-to-one correspondence with the fermions
\cite{Frolov:1993zr,Narayanan:1993zzh,Furman:1994ky,Brower:2012vk}.}

This paper is organized as follows.
In Sec.~\ref{PVsec} we present the PV fields
that accompany the staggered fermions in our simulations.
We  also review the idea that massive fields can be considered
as generating an effective action for the gauge field.
Section~\ref{results} gives an account of our experimentation
with the PV fields, and in Sec.~\ref{conc} we give our conclusions.

\section{\label{PVsec} Pauli--Villars fields}

\subsection{\label{action} Lattice definitions}
In this paper we use a gauge action that includes
fundamental and adjoint plaquette terms
with couplings $\beta\equiv\beta_F$ and $\beta_A$, respectively,
related by $\beta_A / \beta_F = -0.25$ \cite{Cheng:2011ic}.
The gauge links in the staggered operator $D$ are nHYP-smeared links~\cite{Hasenfratz:2001hp, Hasenfratz:2007rf}
with smearing parameters $\alpha = (0.5, 0.5, 0.4)$.
This lattice action has been used in several studies
of 8- and 12-flavor systems, including explorations
of the phase diagram~\cite{Cheng:2011ic, Hasenfratz:2013uha},
the discrete $\beta$ function~\cite{Hasenfratz:2014rna,Hasenfratz:2016dou},
the scale-dependent mass anomalous dimension~\cite{Cheng:2013eu},
and the large-scale spectrum studies
of the LSD collaboration \cite{Appelquist:2016viq,Appelquist:2018yqe}.

The finite volume gradient-flow coupling is defined by
\cite{Narayanan:2006rf,Luscher:2009eq,Luscher:2010iy,Fodor:2012td}
\be
\label{GF}
\gGF(t;L) = \frac{128\pi^2}{3(N_c^2-1)(1+\delta)}\,
\svev{t^2 E(t)} \ ,
\ee
where $N_c=3$ and the term $1/(1+\delta)$ corrects for the gauge zero modes
due to periodic boundary conditions \cite{Fodor:2012td}.
We evaluate $\gGF$  at flow time fixed by $\sqrt{8t}= cL$;
throughout this paper we use $c = 0.45$.

In \Eq{GF}, $E(t)$ is the (flowed) energy density.
Different operators and flows,
like the plaquette and clover operators, or Wilson and Symanzik flows,
give slightly different $\gGF$, but
for $\sqrt{8t}=0.45\,L$ this variation is usually mild.
In the following we define $\gGF$ using Wilson flow and the clover operator.
Since we use the same flow and operator throughout this work,
the specific choice should not affect the qualitative results.

We define the staggered lattice Dirac operator as
\be
\label{D}
M = D + \mf \ ,
\ee
where $D$ is the single-component, massless, anti-hermitian
staggered operator.
The fermion action constructed using this operator
gives rise to $N_f=4$ continuum flavors when the bare gauge coupling is tuned
to the Gaussian fixed point $g_0^2=0$ (equivalently, $\beta\to\infty$),
provided that the fermion mass $\mf$ is accordingly tuned to zero
in lattice units.

The squared determinant of $M$ can be represented using
a single pseudofermion field $\phi$,
\bea
(\Det M)^2 &=&  \Det(M^\dagger M ) \nonumber\\
&=& \int d\phi\, d\phi^\dagger
\exp\left(-\phi^\dagger \frac{1}{M^\dagger M} \phi \right) .
\label{NF8}
\eea
The path integral that includes $(\Det M)^2$ gives $N_f=8$ flavors
in the continuum limit.  By restricting $M^\dagger M$ to the even sites only,
one obtains $N_f=4$ flavors in the continuum limit.

Pauli--Villars fields have the opposite statistics.
The bosonic inverse determinant is represented similarly as
\bea
\label{PV}
\left[\Det(\MPV^\dagger \MPV)\right]^{-1}
&=&  \nonumber\\
 \int d\Phi\, d\Phi^\dagger &&\hspace{-2.5ex}
\exp\left(-\Phi^\dagger  \MPV^\dagger \MPV \Phi \right),
\eea
where we define
\be
\label{MPV}
\MPV = D + \mPV \ .
\ee
Unlike the pseudofermion action, the PV action is local.
As above, we restrict $\MPV^\dagger \MPV$ to the even sites only.
Including $\NPV$ copies of such PV fields in Eq.~\ref{PV} would amount to $4\NPV$ bosonic Dirac fields in the continuum limit
if we were to tune the PV mass to zero.  We stress, however,
that the PV mass will always be held fixed in lattice units,
in order to decouple the PV fields in the continuum limit.

\subsection{\label{heavy} Integrating out heavy matter fields}

Matter fields with mass of the order of the cutoff
decouple from the infrared dynamics,
but they generate an effective gauge action, affecting the bare gauge coupling.
$N_s$ staggered fermions give  an induced contribution to the gauge action
\bea
e^{-\Sind[U]} &=& \int \prod_{i=1}^{N_s} d\psi_i\, d\bar\psi_i\,
\exp\left[-\sum_{i=1}^{N_s} \bar\psi_i (D+\mf) \psi_i\right]\nonumber\\[3pt]
&=& \Det(D+\mf)^{N_s}
= e^{N_s \Tr \log (D+\mf)} \ .
\label{Zeff}
\eea
The full effective action for the gauge field is thus
\be
\Seff[U]=S_{\text{gauge}}[U]+\Sind[U].
\ee
For thin (unsmeared) links, an expansion in powers of $1/\mf$ gives
\be
\label{Seff}
\Sind = -N_s  \sum_\ell \frac{(-1)^{\ell/2}}{\ell(2a\mf)^\ell}
\sum_{x}\sum_{\mathcal{C}_\ell}  {\cal E}_{\mathcal{C}_\ell} \Tr U_{\mathcal{C}_\ell} \ ,
\ee
where the sum over $\mathcal{C}_\ell$ is a sum over all closed loops
of length $\ell$ originating from lattice site $x$,
and ${\cal E}_{\mathcal{C}_\ell}= \frac{1}{4} \Tr (\gamma_{\mu_1} \cdots \gamma_{\mu_\ell})$ is a sign factor
that depends on the geometry of the loop.

For heavy staggered fermions interacting with thin (unsmeared) gauge links
the leading term in $\Sind$ is a plaquette term with coefficient
\be
\label{bind}
\beta^{(p)}_\text{ind} = \frac{N_s}{(2a\mf)^4} \ .
\ee
The coefficient of $F_{\mu\nu}^2$ from all the loops
(representing the leading term in an expansion in the lattice spacing)
can be expressed as a new inverse bare gauge coupling
$\beta_\text{eff}=\beta + \beta_\text{ind}$.
Smeared link actions generate a smeared plaquette at leading order that also corresponds to an induced action characterized by a shift in $\beta$.

Ref.~\cite{Hasenfratz:1993az} investigated the finite temperature
phase transition of various systems (with thin links) and concluded
that fermions with $a\mf \gtrsim 0.1$ are well described by
an effective gauge action.
The possibility of replacing the entire gauge action
with an effective action generated by fermions
was explored in \cite{Hasenfratz:1992jv}, while \cite{Kazakov:1992ym} considered heavy scalar fields in the adjoint representation.

If we replace the fermions by PV bosons,
the only  change is that the overall sign on the right-hand side
of \Eq{Seff} is flipped.  For equal masses $\mPV=\mf$, the PV bosons
exactly cancel out the $\Sind$ generated by $N_s=\NPV$ staggered fields.
Similarly, in continuum perturbation theory, and for scales $\m\gg \mPV=\mf$,
the contribution of the PV bosons to the one- and two-loop beta function
would be exactly opposite to that of the fermions.

In the next section, we turn to a numerical study of the effect
of PV bosons with masses $a\mPV=\mathcal{O}(1)$.

\section{\label{results} Numerical simulations with Pauli--Villars fields}

We have explored the effect of the PV fields
in the SU(3) gauge theory with $N_f=12$ massless  fermions,
using a staggered action with nHYP smearing (see Sec.~\ref{action}).
This theory exhibits a first-order transition into the S4 phase
\cite{Cheng:2011ic,Hasenfratz:2013uha,Aubin:2015dgk,Kotov:2021mgp},
which persists for all the combinations of PV fields that we have tried out.

\begin{figure}[t]
\vspace*{-1ex}
\includegraphics[width=0.99\columnwidth]{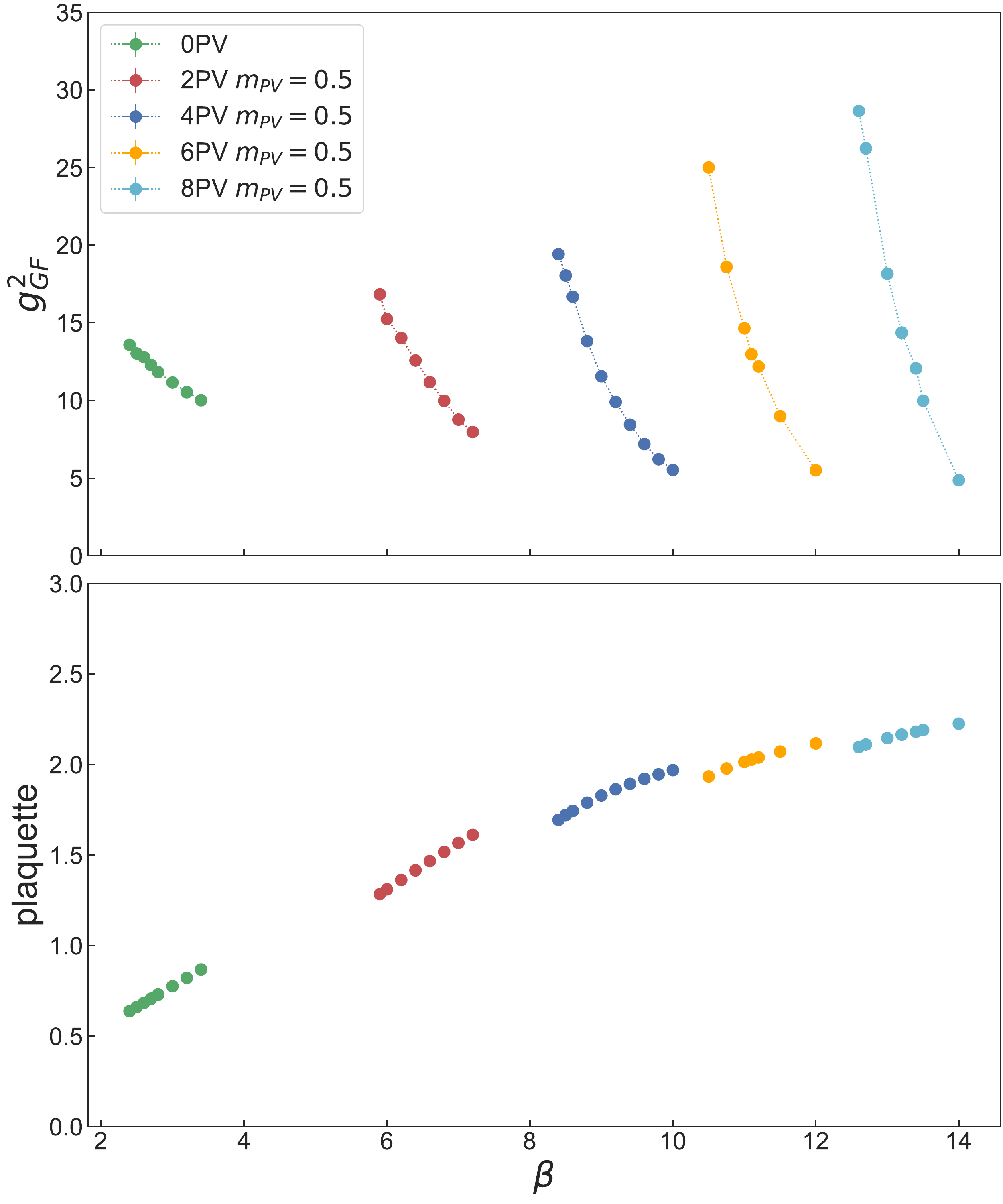}
\vspace*{-2ex}
\floatcaption{Nf12-beta-nhyp-PVs2}%
{
The effect of increasing the number of  PV fields.
The top panel shows the gradient-flow coupling $\gGF$
and the bottom panel the plaquette \vs\ the bare coupling $\beta$
for the SU(3) theory with $N_f=12$ fermions in the chiral limit.
All systems exhibit a first-order transition to the $S^4$ phase.
As before, the leftmost point in each set is a little to the right
of the phase transition; data points inside the S4 phase are not shown.
}
 \end{figure}

In Fig.~\ref{Nf12-beta-nhyp-PVs2} we show the effect of a growing number
of PV fields, always keeping $a\mPV=0.5$.
We consider $\NPV=0,2,4,6,$ and 8.
The lattice volume is $8^4$.
The top panel of Fig.~\ref{Nf12-beta-nhyp-PVs2} shows $\gGF$
as a function of the inverse bare gauge coupling $\beta$.
A striking feature is how much
fixed physics, defined by a given value of $\gGF$, gets pushed back towards larger $\beta$. This is  thanks to the anti-screening effect of the PV fields.
For example, let us consider $\gGF=11$.
Without PV fields, this value is achieved for $\beta\simeq 2.8$
(see Table~\ref{table:perf}).
With $\NPV=2$, $a\mPV=0.5$, the same renormalized coupling is attained for $\beta\simeq 6.8$.
With each further addition of 2 to $\NPV$, the value of $\beta$ shifts by about 2.2.
This behavior is qualitatively consistent with the predictions
of the hopping-parameter expansion, see Sec.~\ref{heavy}.

\begin{table}[t]
\centering
\begin{ruledtabular}
\begin{tabular}{crlcrl}
 $\NPV$	&  $\beta$ &$\gGF$ &   $N_{\textrm{step}}$ & $N_{\textrm{CG}}$ & $|\delta H| $\\
\hline
 0  &  2.8   &    10.87(6)   &   15 &     1190     & 0.45  \\
 2  &  6.8   &    10.89(9)   &   12 &      400     & 0.062 \\
 4  &  9.0   &    10.58(8)   &   12 &      365     & 0.033 \\
 6  & 11.2   &    10.87(30)  &   12 &      353     & 0.033 \\
 8  & 13.4   &    10.87(100) &   12 &      359     & 0.034 \\
\end{tabular}
\end{ruledtabular}
\caption{
Performance of simulations with different PV content that yield $\gGF\simeq11$,
on volume $8^4$ with $N_f=12$ staggered fermions.
The fermion mass is $a\mf=0$ while $a\mPV=0.5$.  Trajectories are of unit length.
$N_{\textrm{step}}$ is the number of MD steps per trajectory,
$N_{\textrm{CG}}$ is the average number of CG iterations
for the pseudofermion inversions, and $|\delta H|$ is the average error
in the MD energy.}
\label{table:perf}
\end{table}

The bottom panel of Fig.~\ref{Nf12-beta-nhyp-PVs2} shows that increasing the number of PV fields
while keeping the renormalized coupling fixed is accompanied by
increasing plaquette values.
This demonstrates that the short-distance fluctuations
of the gauge field are damped. The system without PV bosons has plaquette expectation value around 0.6 (out of 3) just above the bulk transition. Such rough gauge fields would not be considered acceptable in QCD simulations.

Another  striking feature seen in the top panel of Fig.~\ref{Nf12-beta-nhyp-PVs2}
is the larger range of renormalized gauge couplings made accessible when PV fields are added.  As in Fig.~\ref{gsq-beta-nhyp-noPV},
the leftmost point in each data set is just to the right of
the phase transition into the S4 phase.  As we increase $\NPV$ from left to right,
the corresponding maximal values of $\gGF$ are approximately~14, 17, 20, 25,
and~29.%
\footnote{We have identified the location of the transition to the $S^4$ phase
to within about 0.1 in $\beta$. The quoted maximal values of $\gGF$
have  statistical and systematic errors of about  0.3 and~1.0, respectively.}
The largest renormalized coupling attainable with
$\NPV=8$ is about twice that without any PV fields.

\begin{figure}[t]
\vspace*{-1ex}
\includegraphics[width=0.99\columnwidth]{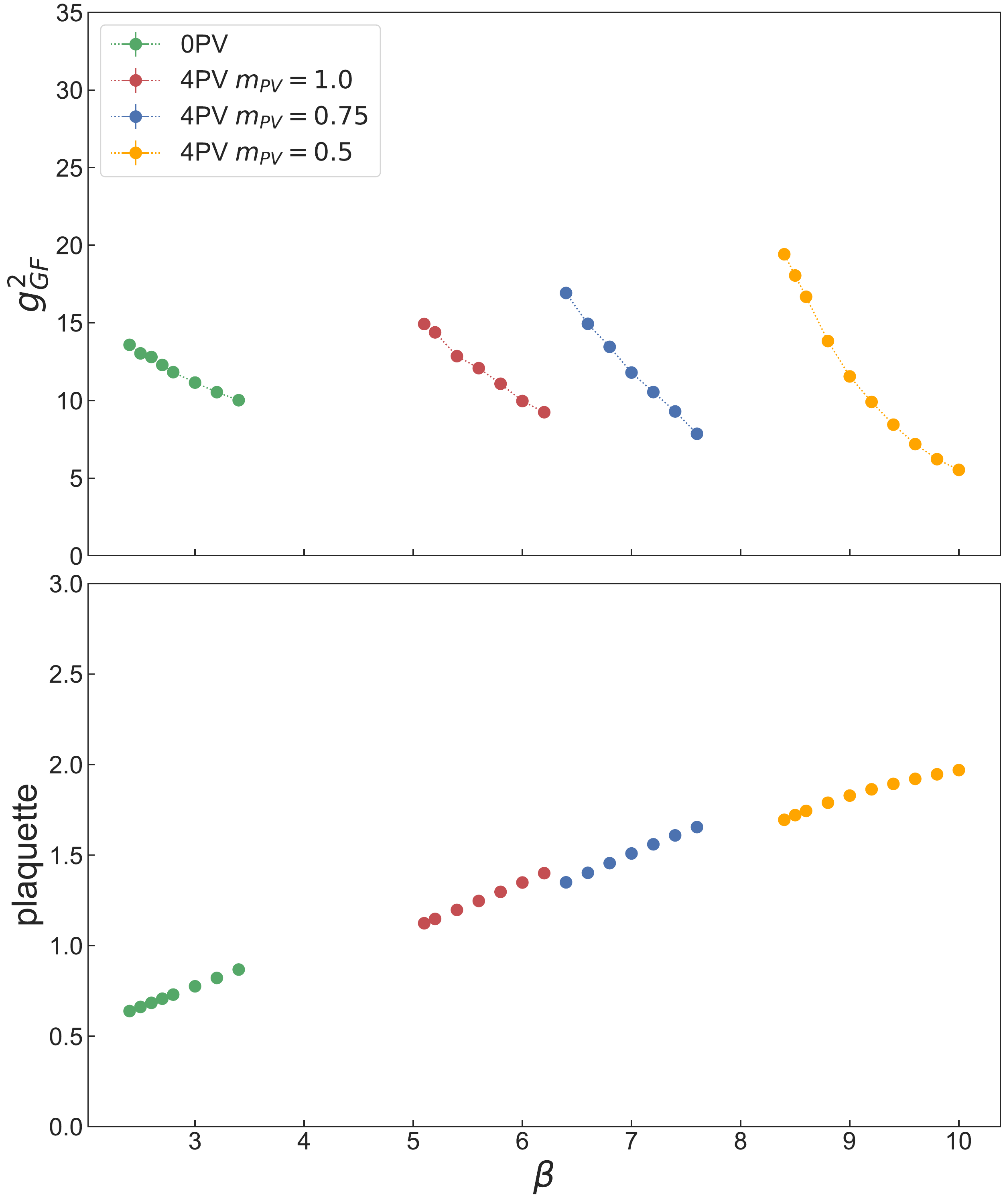}
\vspace*{-2ex}
\floatcaption{Nf12-beta-nhyp-PVs}%
{
Similar to Fig.~\ref{Nf12-beta-nhyp-PVs2},
except the number of PV fields, where present, is fixed at $\NPV=4$,
with decreasing values of $\mPV$.
Again all systems exhibit a first-order transition to an $S^4$ phase,
and the leftmost point in each set is a little to the right
of the phase transition.
}
\end{figure}

Fig.~\ref{Nf12-beta-nhyp-PVs} is similar to Fig.~\ref{Nf12-beta-nhyp-PVs2},
but this time we keep the number of PV fields fixed at $\NPV=4$
and we consider three values of the PV mass, from large to small:
$a\mPV=1.0$, 0.75, and~0.5.  As expected, the effect of the PV fields
increases with decreasing mass.
Comparing Figs.~\ref{Nf12-beta-nhyp-PVs} and~\ref{Nf12-beta-nhyp-PVs2}
we deduce, for example,
that $\NPV=4$  with mass $a\mPV=0.75$ has roughly the effect of
$\NPV=3$ with $\mPV=0.5$. If parameters of the simulation require it,
one can always raise the PV mass to $a\mPV=1$, and then increase $\NPV$
till the desired anti-screening effect is achieved.
(There is no reason to consider heavier PV bosons,
since $a\mPV=1$ already sets the PV mass equal to the cutoff.)
We stress again that the PV mass should be
held fixed in lattice units when taking the continuum limit.

The reader may wonder if  PV bosons with mass $a\mPV=0.5$ are heavy enough
to decouple.  First, we have found that the mass of the PV pion---the
ghost pion made of two PV bosons---is roughly equal to $2\mPV$.
This implies that the couplings between remote sites in $\Sind$
decrease exponentially with a decay rate $\sim 2\mPV$.

We can also examine this question from the point of view
of the infrared scales.
In systems where chiral symmetry is not broken,
the only dimensionful parameters, besides $m_{PV}$,
are the fermion masses (denoted generically as $\mf$)
and the lattice volume.
Fermion bound states are expected to scale with $\mf$,
so as  long as $m_{PV} \gg  \mf$ and   $m_{PV} \gg 1/L$,
the PV bosons decouple. This applies both in mass-deformed conformal systems
and in QCD-like systems at high temperature, where there is no condensate.
In particular, if simulations are performed in the chiral  limit,
the only condition is $m_{PV} L \gg 1$.
Where chiral symmetry is broken,
an additional condition $m_{PV} \gg \Lambda_{QCD}$ has to be imposed.
In practice one can check that $m_{PV} \gg M_H$
where $M_H$ is the heaviest physically accessible hadron of the system.

\begin{figure}[t]
\vspace*{-1ex}
\includegraphics[width=0.99\columnwidth]{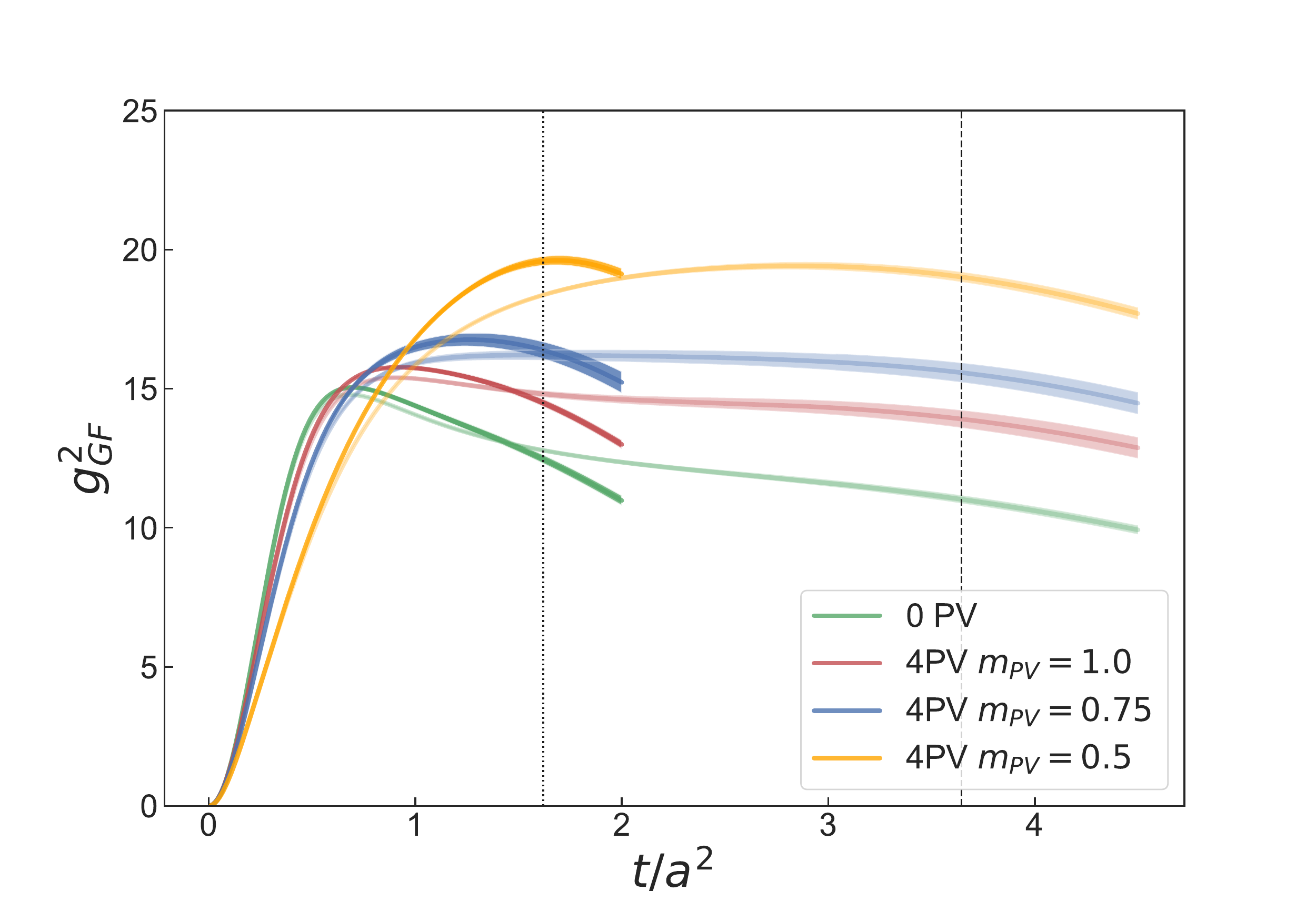}
\vspace*{-2ex}
\floatcaption{maxg2-flow-4PV}%
{
The gradient flow coupling $\gGF(t;L)$ as a function of the flow time $t/a^2$
at couplings $\beta$ just above the respective bulk transitions
(see Fig.~\ref{Nf12-beta-nhyp-PVs}). We show data with no PV fields
and with 4 PV  bosons with decreasing mass $\mPV$, on volume $8^4$ (short curves) and $12^4$ (long curves).
The vertical lines correspond to flow time at $c=0.45$ for the two volumes.
}
\end{figure}


The flow-time evolution of $\gGF(t;L)$ is always characterized by
three regions.  It starts with an initial rise at $t/a^2 \lesssim \mathcal{O}(1)$ that depends strongly on the specific action,
flow, and operator choice.  This is followed by a physical region,
and finally the flow saturates due to the finite volume.
In Fig.~\ref{maxg2-flow-4PV} we show the $t$ dependence of $\gGF(t;L)$
for $\beta$ just above the S4 transition, as shown
in Fig.~\ref{Nf12-beta-nhyp-PVs}.  In the absence of any PV fields
the ``knee'' where $\gGF$ turns from the initial short-distance rise
to the physical (and, here, decreasing) part occurs at $t/a^2 \simeq 0.6$.
The inclusion of PV fields induces a short-distance gauge interaction
that affects the small-$t$ behavior.
In the flows with 4 PV fields the turning point
between the two regions moves to the right as $\mPV$ is lowered,
while the maximum of $\gGF$ grows.

For $c=0.45$ and $L=8$ we read off the gradient-flow coupling
at $t/a^2 \simeq 1.62$.  This small volume obviously does not leave much room
for a physical region.  In Fig.~\ref{maxg2-flow-4PV} we also show
the flows for $L=12$.  Here the gradient-flow coupling is determined
at $t/a^2 \simeq 3.65$, which is now well beyond the turning point
of the flows for all values of $\mPV$.
Both the maximum of $\gGF(t;L)$ and its value at $\sqrt{8t}= 0.45\,L$
do not change much with the volume.%
\footnote{If, as in Fig.~\ref{Nf12-beta-nhyp-PVs2}, we hold $\mPV$ fixed
and vary $\NPV$ we observe a similar trend: again the turning point
moves to the right with increasing $\NPV$
and the maximum of $\gGF$ grows.}

Finally, we turn to the cost of the PV fields.
As long as the number of PV bosons is not extremely large,
adding them to the simulation is inexpensive.
The PV action~(\ref{PV}) requires only one inversion (with large mass)
at the beginning of an HMC trajectory to generate each PV field $\Phi$.
During the molecular dynamics evolution of the gauge field, calculating the force due to the PV action requires mutiplication with the
Dirac operator, but no inversion.

As a matter of fact, the PV fields can even make the simulations less expensive.
In Table~\ref{table:perf} we report the performance of simulations
with different numbers of PV fields with $\mPV=0.5$.%
\footnote{Chiral symmetry breaking was not observed in any of our data sets.
This allowed us to work directly at $a\mf=0$.}
As $\NPV$ is increased from zero,
we change $\beta$ to keep $\gGF\simeq11$ fixed
(cf.~Fig.~\ref{Nf12-beta-nhyp-PVs2}).
We observe decreases in (1) the number of conjugate-gradient (CG) iterations
and in (2) the average error $|\delta H|$ in the molecular dynamics (MD) energy
due to large forces in the equations of motion.
Sharp decreases are seen in going from $\NPV=0$ to~2
despite an increased step size $d\tau=1/N_{\textrm{step}}$.%
\footnote{Adding more than 4 PV fields seems to have no further effect.}
A decrease in $|\delta H|$ brings about improved acceptance
and should allow yet larger values of $d\tau$,
further decreasing the computational cost.

\section{\label{conc} Discussion}

In this paper we investigated the possibility of countering
the effective action generated by fermions at short distances
by including heavy Pauli--Villars fields.
The effective action generated by PV bosons has the opposite sign to that generated by fermions;
while fermions screen the gauge coupling, PV bosons anti-screen.
As long as their mass $a\mPV$ is kept
at ${\cal O}(1)$, the PV bosons decouple from the infrared dynamics.

 Pauli--Villars fields are part of standard domain-wall fermion actions.
Indeed, domain-wall fermions are generally known for their small cutoff effects.  In this paper, we considered
the use of a more general set of PV fields with staggered fermions.
Similarly, one can add heavy PV bosons to a theory with Wilson fermions.
We would always use the same lattice Dirac operator for the
unphysical PV bosons as for the fermions (except of course with a heavy mass $\mPV$),
anticipating that this is the optimal way to reduce the cutoff effects
of the fermions. One can also try using a different PV action; the PV fields would simply
induce a different gauge action.

Increasing the number of heavy PV bosons (even beyond the number of fermions)  enlarges the  induced gauge action. As a consequence, the same physical regime is reached at smaller bare  gauge coupling (\ie  larger $\beta$) where the gauge fields are smoother.
 The ability to work at small bare coupling
while maintaining the desired renormalized coupling in the infrared
means that cutoff effects are reduced.  In a way,
introducing many PV bosons brings the system closer to a perfect action.

Heavy PV bosons have an additional positive effect on systems that exhibit  bulk first order phase transitions in  strong coupling.
These phase transitions are typically caused by large ultraviolet fluctuations and are not physical. Nevertheless they limit the range of  possible renormalized couplings. While the  additional PV bosons in our test study with $N_f=12$ flavors did not remove the bulk transition, they shifted the simulations to smoother gauge fields and opened up the range of accessible couplings. In our experiments we found  about a factor of two increase in the maximal $\gGF$.

Adding several PV fields to the simulation is inexpensive.
In fact, the PV fields made our simulations less expensive:
Since the gauge fields are smoother, the CG inversions needed
for the fermion force converged significantly faster.
Moreover, the smaller MD forces resulted in improved acceptance
even with a larger step size.

The introduction of PV bosons can be particularly useful
for slowly running systems where, for the renormalized coupling
at the infrared scale to be large,
normally the bare coupling would have to be taken large, too.
PV bosons allow for independent control over the magnitude
of discretization effects in such systems.

Our aim in this paper has been to show that the addition of PV fields
smooths the gauge field and
allows the gradient flow to reach very strong couplings.
We have demonstrated these claims using $N_f=12$ flavor staggered fermions.
Any conclusions to be drawn from the data presented will, as usual,
depend on the results of taking large-volume and continuum limits.
The question of whether the $N_f=12$ theory is ultimately confining
\cite{Fodor:2011tu,Fodor:2016zil,Fodor:2017gtj,Fodor:2019ypi} or conformal \cite{Hasenfratz:2016dou,Hasenfratz:2019dpr} has been the subject of extensive study.
If the downward trend of the flows in the physical region (Fig.~\ref{maxg2-flow-4PV}) survives all relevant limits, this would signal the presence of an infrared fixed point
at some smaller value of $g^2_{GF}$.

PV bosons can find their use in QCD, too.
For given computer resources, the only way to increase
the physical volume is by increasing the lattice spacing.
This comes at the price of growing discretization errors.
Including PV bosons might allow for simulations with larger
lattice spacings and hence with larger physical volumes,
while still keeping the discretization effects under control.
At the same time simulations with PV bosons could
be computationally faster as well.

\vspace{2ex}
\section*{Acknowledgments}
Computations for this work were carried out
in part on facilities of the USQCD Collaboration, which are funded by the Office of Science of the U.S.~Department of Energy, the RMACC Summit supercomputer, which is supported by the National Science Foundation (awards No.~ACI-1532235 and No.~ACI-1532236),
the University of Colorado Boulder, and Colorado State University, and on  the University of Colorado Boulder HEP Beowulf cluster. We thank the University of Colorado for providing the facilities essential for the completion of this work.
A.H. acknowledges support by DOE grant DE-SC0010005.
The work of B.S. and Y.S. was supported by the Israel Science Foundation under grant No.~491/17.

\bibliography{PVpaper}

\end{document}